\documentclass[twocolumn,preprintnumbers,superscriptaddress,nofootinbib,aps,prd,floatfix]{revtex4}
\pdfoutput=1
\usepackage{datetime}
\usepackage{color}

\usepackage{footmisc,multirow}
\usepackage{subfigure}
\usepackage{amsmath}
\usepackage{natbib}
\usepackage{xspace}
\usepackage{graphicx}
\usepackage{booktabs,hhline}
\usepackage{array}
\usepackage{graphicx}
\usepackage[utf8]{inputenc}

\DeclareMathAlphabet\mathbfcal{OMS}{cmsy}{b}{n}

\newcommand{\nc}{\newcommand}

\nc{\newsection}[1]{\section{#1}\setcounter{equation}{0}}
\nc{\newappendix}[1]{\section*{#1}\setcounter{equation}{0}}
\nc{\scm}{\scriptscriptstyle\text}
\nc{\f}{\frac}

\nc{\bea}{\begin{eqnarray}}   \nc{\eea}{\end{eqnarray}}
\nc{\baa}{\begin{array}}      \nc{\eaa}{\end{array}}
\nc{\bit}{\begin{itemize}}    \nc{\eit}{\end{itemize}}
\nc{\ben}{\begin{enumerate}}  \nc{\een}{\end{enumerate}}
\nc{\bce}{\begin{center}}     \nc{\ece}{\end{center}}
\nc{\bfl}{\begin{flushright}} \nc{\efl}{\end{flushright}}
\nc{\btb}{\begin{tabular}}    \nc{\etb}{\end{tabular}}
\nc{\eps}{\varepsilon}
\nc{\vp}{\varphi}
\nc{\tvp}{\widetilde{\varphi}}
\nc{\D}{\mbox{$\not\!\!D$}}
\nc{\Db}{\mbox{${\raisebox{2mm}{\boldmath ${}^\leftarrow$}\hspace{-4mm} D}$}}
\nc{\Dfb}{\mbox{$\raisebox{2mm}{\boldmath ${}^\leftrightarrow$}\hspace{-4mm} D$}}
\nc{\vpj }{\mbox{${\vp^\dag i\,\raisebox{2mm}{\boldmath ${}^\leftrightarrow$}\hspace{-4mm} D_\mu\,\vp}$}}
\nc{\vpjt}{\mbox{${\vp^\dag i\,\raisebox{2mm}{\boldmath ${}^\leftrightarrow$}\hspace{-4mm} D_\mu^{\,I}\,\vp}$}}
\newcommand{\be}{\begin{eqnarray*}}
\newcommand{\ee}{\end{eqnarray*}}

\newcommand{\bee}{\begin{eqnarray}}
\newcommand{\eee}{\end{eqnarray}}
\newcommand{\beeq}{\begin{equation}}
\newcommand{\eeeq}{\end{equation}}

\newcommand{\gev}{{\text{GeV}}\xspace}
\newcommand{\tev}{{\text{TeV}}\xspace}
\newcommand{\ifb}{\ensuremath{\text{fb}^{-1}}\xspace}
\newcommand{\iab}{\ensuremath{\text{ab}^{-1}}\xspace}
\newcommand{\sqrts}{\ensuremath{\sqrt{s}}\xspace}

\newcommand{\lag}[1]{\ensuremath{\mathcal{L}_\text{#1}}}
\newcommand{\summ}[1]{\ensuremath{\sum_{\substack{#1}} }}
\newcommand{\co}[2][]{\ensuremath{C_{#2}^{#1}}}
\newcommand{\bco}[2][]{\ensuremath{\bar{C}_{#2}^{#1}}}
\newcommand{\op}[2][]{\ensuremath{\mathcal{O}_{#2}^{#1}}}

\preprint{MCNET-16-26}

\begin{document}

\title{Giving top quark effective operators a boost}

\begin{abstract}
  We investigate the prospects to systematically improve generic
  effective field theory-based searches for new physics in the top
  sector during LHC run 2 as well as the high luminosity phase.  In
  particular, we assess the benefits of high momentum transfer final
  states on top EFT-fit as a function of systematic uncertainties in
  comparison with sensitivity expected from fully-resolved analyses
  focusing on $t\bar t$ production. We find that constraints are
  typically driven by fully-resolved selections, while boosted top
  quarks can serve to break degeneracies in the global fit. This
  demystifies and clarifies the importance of high momentum transfer
  final states for global fits to new interactions in the top sector
  from direct measurements.
\end{abstract}

\author{Christoph Englert} \email{christoph.englert@glasgow.ac.uk}
\affiliation{SUPA, School of Physics and Astronomy, University of
  Glasgow,\\Glasgow, G12 8QQ, United Kingdom\\[0.2cm]}

\author{Liam Moore} \email{l.moore.1@research.gla.ac.uk}
\affiliation{SUPA, School of Physics and Astronomy, University of
  Glasgow,\\Glasgow, G12 8QQ, United Kingdom\\[0.2cm]}
\affiliation{Theory Division, CERN, 1211 Geneva 23, Switzerland\\[0.2cm]}

\author{Karl Nordstr\"om} \email{k.nordstrom.1@research.gla.ac.uk}
\affiliation{SUPA, School of Physics and Astronomy, University of
  Glasgow,\\Glasgow, G12 8QQ, United Kingdom\\[0.2cm]}
\affiliation{Institut f\"ur Theoretische Physik, Universit\"at Heidelberg,\\Philosophenweg 16, 69120 Heidelberg, Germany\\[0.2cm]}
  
\author{Michael Russell} \email{m.russell.2@research.gla.ac.uk}
\affiliation{SUPA, School of Physics and Astronomy, University of
  Glasgow,\\Glasgow, G12 8QQ, United Kingdom\\[0.2cm]}
\affiliation{Institut f\"ur Theoretische Physik, Universit\"at Heidelberg,\\Philosophenweg 16, 69120 Heidelberg, Germany\\[0.2cm]}

\maketitle

\section{Introduction}
\label{sec:intro}
Final states associated with top quarks are produced in abundance
at the Large Hadron Collider (LHC). Top quark pair
production in particular, with a cross section of around
900~pb~\cite{Moch:2008qy,Aliev:2010zk,Czakon:2013goa,Czakon:2016ckf}, will enable us to perform precise spectroscopy of the
top sector at the LHC. This plays an important role in paving the way
to a better understanding of particle physics beyond the Standard
Model (SM).  In fact, since the top quark is the heaviest particle in the
SM it typically assumes a central role in concrete beyond the SM (BSM) scenarios,
ranging from composite Higgs to supersymmetric theories. Most of these
theories are characterised by additional propagating degrees of
freedom, some of which fall inside the kinetic coverage of the LHC.
There is a significant effort to search for these exotic states. Unfortunately, none of these searches have provided a
conclusive hint for physics beyond the SM so far.

If there is a mass gap between the electroweak scale $v$ and the scale of
the new physics, one can view the Standard Model as the leading term in
an effective Lagrangian, where all non-standard couplings of SM
particles to new degrees of freedom are integrated out and are left encoded in higher-dimensional
operators, i.e. operators of dimension $D > 4$:
\begin{equation}
\lag{eff} = \lag{SM}+\summ{i}\frac{\co[(6)]{i}\op[(6)]{i}}{\Lambda^2}+\ldots
\label{eqn:eftexpn}
\end{equation}
As is convention, we normalise the operators such that $\Lambda$, which can be viewed as a generic scale for the supposed heavy degrees of freedom, has been displayed explicitly, so that the effective new physics couplings \co[(6)]{i} (Wilson coefficients) are dimensionless. The ellipsis denotes operators of dimension $D > 6$. It is typically assumed that there is an adequate gap between $v$ and $\Lambda$ such that this expansion can be truncated at $D = 6$, although this may not be the case. 

The structure of the operators \op[(6)]{i} is dictated entirely by the field content and symmetry restrictions of the Standard Model, and several equivalent bases for \op[(6)]{i} exist already in the literature. Once a basis has been chosen, all that remains is to compute the effects of \op[(6)]{i} on a given observable, and obtain the allowed regions of its corresponding coefficient \co[(6)]{i}. Any measurement of a non-zero coefficient is thus evidence of physics beyond the Standard Model.

The simplicity and generality of the effective field theory approach to LHC measurements is evident from the vast literature already existing on the subject, in which several directions of study have been pursued.  For instance, theoretical improvements of the framework itself have been provided in~\cite{Hagiwara:1993ck,Degrande:2014tta,Jenkins:2013zja,Jenkins:2013wua,Alonso:2013hga,Ghezzi:2015vva,Zhang:2016omx,Franzosi:2015osa,Berthier:2015oma,Elias-Miro:2014eia}, and approaches to confront the multi-dimensional parameter spaces of Wilson coefficients relevant to a given class of observable with the plethora of presently available and expected future measurements are developed in~\cite{Corbett:2012ja,Corbett:2015mqf,Butter:2016cvz,Buckley:2015nca,Buckley:2015lku,Englert:2015hrx,Falkowski:2015jaa,Efrati:2015eaa,Falkowski:2014tna,Falkowski:2015krw,Castro:2016jjv,Cirigliano:2016nyn,Rosello:2015sck,deBlas:2015aea}. 

The top quark sector is one of the pillars of the analysis programme pursued by the ATLAS and CMS experiments at the LHC, and the diverse range of measurements published during run 1 have been used to constrain new top couplings within the SMEFT framework~\cite{Buckley:2015nca,Buckley:2015lku,Rosello:2015sck}. No deviations have been found so far (in particular in resonant $t\bar t$ searches e.g.~\cite{Aad:2012em,Aad:2012ans,Chatrchyan:2012yca}), but the constraints available from the small integrated luminosity of run 1 are rather weak. This is a setback for the EFT approach, because weak constraints on $\co{i}/\Lambda^2$ can correspond to mass scales $\Lambda$ that are resolved by the measurements used in the fit, thus invalidating the {\emph{perturbative}} expansion\footnote{An identical interpretation is that weakly coupled UV completions are left unconstrained after matching.} of Eq.~\eqref{eqn:eftexpn} in the first place~\cite{Buckley:2015lku,Englert:2014cva,Contino:2016jqw}

With the increase in luminosity forecast for run 2, as well as
the high luminosity-phase of the LHC (or even a hypothetical 100 TeV
collider), there will undoubtedly be a significant improvement of the
currently rather loose constraints. This improvement
will depend on the relative importance of particular top channels and
the impact of their associated systematic uncertainties on the limit
setting. We expect deviations from the SM to be most pronounced at
large momentum transfers. However, in these regions both the
theoretical modelling as well as the experimental measurements tend to
be less reliable than at low momentum transfers. The small cross
sections at large transverse momenta can be mitigated by employing
efficient top reconstruction techniques in these particular phase
space regions and these reconstruction approaches are subject to
qualitatively different systematics compared to fully resolved top
final state analyses. This holds in particular when new physics is
present.

In this paper we show how the combination of these effects influences
the potential improvement of a global top quark fit using the example
of top quark pair production and discuss how improvements in boosted
and fully resolved analyses can affect the global sensitivity to new
physics in top final states. We organise this work as follows. We
first introduce the relevant effective theory and fitting components which are
relevant for our analysis in Sec.~\ref{sec:elements}. In Sec.~\ref{sec:improve} we discuss the improvements made on the limits of the coefficients as obtained from the boosted analysis, highlighting the interplay between various sources of experimental uncertainty, and trace the impact of theory uncertainties on our fit. We conclude in Sec.~\ref{sec:conc} with a discussion of how our EFT results may then be interpreted within the parameter space of UV completions, and discuss future directions.

\begin{figure*}[t!]
  \begin{minipage}[l]{0.67\textwidth}
    \includegraphics[width=\textwidth]{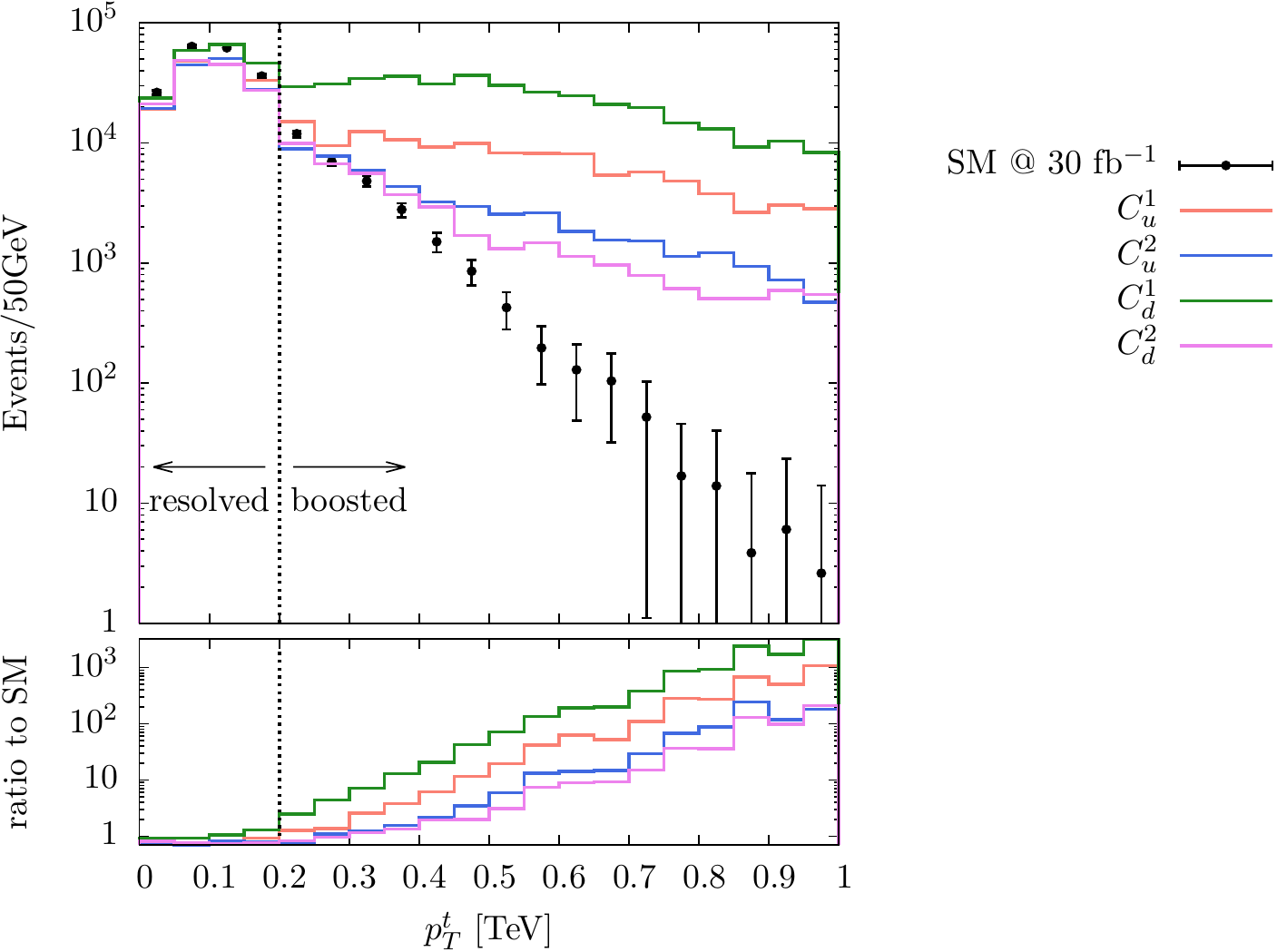}
  \end{minipage} 
  \hspace{-3.2cm}
  \begin{minipage}[r]{0.45\textwidth}
  \vspace{3.cm}
    \caption{Transverse momentum distributions for the reconstructed hadronic top quark candidate. The bars represent 30~\ifb of pseudodata with \sqrts = 13 \tev constructed with the SM-only hypothesis, while the shaded curves include the effects of four-quark operators with Wilson coefficients $C_i = 10 $ TeV$^{-2}$ for illustration. Details of the top quark reconstruction are described in the text. \label{fig:boosted_pt}}
  \end{minipage}
\end{figure*}

\section{Elements of top fitting}
\label{sec:elements}

\subsection{Model Summary}
\label{sec:model}

We implemented the complete 'Warsaw' Basis~\cite{Grzadkowski:2010es} of the
SMEFT Lagrangian as a general \textsc{FeynRules}~\cite{Alloul:2013bka}
model, in which we adopted these authors' conventions and permitted Wilson
coefficients $C$ to carry flavour indices wherever applicable. 
Herein, we included the minimal set of global parameter and field
redefinitions necessary to restore canonical normalisation and
mass-diagonal states to the Lagrangian in the broken electroweak phase up to terms of
$\mathcal{O}(\Lambda^{-4})$. In the case of strong top pair production this amounts to redefinitions of e.g. the strong gauge coupling 
\begin{equation}
  g_s \to  g_s \left(1 + C_{\Phi G} \, {v^2\over \Lambda^2}  \right) 
\end{equation}
(see e.g.~\cite{Berthier:2015oma} for a detailed discussion), which have no physical consequences for our analysis.

The resulting model file is interfaced via the {\textsc{Ufo}}~\cite{Degrande:2011ua} format to {\textsc{MadEvent}}~\cite{Alwall:2014hca}. At leading order in the SMEFT, the operators that are relevant for top quark pair production at hadron colliders are summarised in Tab.~\ref{tab:ops}.

\setlength\extrarowheight{2.5pt}
\begin{table}[!t]
\begin{tabular}{ll@{\qquad}ll} \hline
\textbf{Coefficient \co{i}} & \textbf{Operator \op{i}} \\  \hline 
\co{G} &  $f_{ABC} G_{\mu}^{A \nu}G_{\nu}^{B \lambda} G_{\lambda}^{C \mu} $ \\
\co{uG} & $ (\bar{q}\sigma^{\mu \nu} T^A u)\tilde \varphi G_{\mu\nu}^{A} $\\
\co[1]{qq} & $ (\bar{q}\gamma_{\mu}q)( \bar{q}\gamma^{\mu}q) $\\
\co[3]{qq} & $ (\bar{q}\gamma_{\mu}\tau^Iq)( \bar{q}\gamma^{\mu}\tau^I q) $ \\
\co{uu} & $ (\bar{u}\gamma_{\mu}u)( \bar{u}\gamma^{\mu} u) $ \\ 
\co[8]{qu} & $ (\bar{q}\gamma_{\mu}T^Aq)( \bar{u}\gamma^{\mu} T^Au) $ \\
\co[8]{qd} & $ (\bar{q}\gamma_{\mu}T^Aq)( \bar{d}\gamma^{\mu} T^Ad) $ \\ 
\co[8]{ud} & $ (\bar{u}\gamma_{\mu}T^Au)( \bar{d}\gamma^{\mu} T^Ad) $ \\ \hline
\end{tabular}
\caption{\label{tab:ops} The operators impacting top pair production considered in this work here.}
\end{table}

The lower six operators in Tab. \ref{tab:ops} contribute via the partonic subprocess $q\bar{q}\to t\bar{t}$, but at the interference level, only through four linear combinations which we denote \co[1,2]{u,d} (see \cite{Buckley:2015lku} for details).

\subsection{Fitting}

The events generated from {\textsc{MadEvent}} which sample the Wilson coefficient space are subsequently showered by {\textsc{Herwig++}}~\cite{Bahr:2008pv,Bellm:2015jjp}, which takes into account initial and final state radiation showering, as well as hadronisation and the underlying event. At this stage, all our predictions are at leading order in the Standard Model EFT. While considerable progress has recently been made in extending the effective Standard Model description of top quark physics to next-to-leading order~\cite{zhang,Degrande:2014tta}, the full description of top quark pair production is incomplete at this order. We take into account higher-order QCD corrections by re-weighting the Standard Model piece of our distributions to the NLO QCD prediction with $K$-factors, as obtained from {\textsc{Mcfm}}~\cite{Campbell:2010ff} and cross-checked with {\textsc{Mc@Nlo}}~\cite{Alwall:2014hca}. Recently, full NNLO results for top quark pair production have become available in~\cite{Czakon:2016ckf,Czakon:2013goa,Moch:2012mk}, we will comment on their potential for improving our results in Sec. \ref{sec:improve}.

We estimate scale uncertainties in the usual way: For the central value of the distributions we choose renormalisation and factorisation scales equal to the top quark mass $\mu_R = \mu_F = m_t$. Then we vary the scales independently over the range $m_t/2 < \mu_{R,F} < 2m_t$. PDF uncertainties are estimated by generating theory observables with the {\textsc{Ct14}}~\cite{Dulat:2015mca}, {\textsc{Mmht14}}~\cite{Harland-Lang:2014zoa} and {\textsc{Nnpdf3.0}}~\cite{Ball:2014uwa} as per the recommendations of the {\textsc{Pdf4Lhc}} working group for LHC run 2~\cite{Butterworth:2015oua}, and we take the full scale+PDF envelope as our theory band. This defines an uncertainty on the differential $K$-factor which we propagate into each observable. We treat theory uncertainties as uncorrelated with experimental systematics and take them to be fixed as a function of luminosity unless stated otherwise.

In order to build the parameter space for the Wilson coefficients \co{i}, instead of calculating coefficients on a multidimensional grid, which suffers from exponential scaling in the number of operators, we use an interpolation-based method, detailed in~\cite{Buckley:2009bj}.
\begin{itemize}
\item We construct a logarithmically random-sampled 6 dimensional parameter space in the operators of Tab. 1. The logarithmic spacing reflects that we want our sampling to be most accurate near to the SM point \{\co{i}\} = 0.
\item We generate our theory predictions and uncertainties, as detailed above, at each point in this space.
\item Once the parameter space has been constructed, we use a polynomial to interpolate between the randomly chosen values of \{\co{i}\}, thus building up a smooth functional form for the change in the prediction for the observables considered with respect to \{\co{i}\}.
\end{itemize} 

Motivated by the functional form of the cross section with respect to the Wilson coefficient
\begin{equation}
\hbox{d}\sigma \sim \hbox{d}\sigma_{\text{SM}}+\co{i}\hbox{d}\sigma_{\text{D6}}+\co[2]{i}\hbox{d}\sigma_{\text{D6}^2}, 
\label{eqn:sigmad6}
\end{equation}
we choose a polynomial dependence on \{\co{i}\} as our response function for a single bin $b$.
\begin{equation}
f_b(\{\co{i}\}) = \alpha_0^b + \summ{i}\beta_i^b\co{i}+\summ{i\leq j}\gamma^b_{i,j}\co{i}\co{j} + \ldots   .
\label{eqn:intpoly}
\end{equation}
This way operators with vanishing interference with the SM amplitude piece can be treated separately and we gain complete analytical control over the fit. The ellipsis in Eq.~\eqref{eqn:intpoly} denotes higher order terms in \{\co{i}\}. Comparing Eqs. (\ref{eqn:sigmad6}) and (\ref{eqn:intpoly}), one would expect a quadratic polynomial to capture the full dependence on \{\co{i}\}. However, when one considers observables such as asymmetries, or distributions normalised to the total cross section, this simple relation is no longer valid. In order to capture the dependence on the coefficients as accurately as possible, we use a fourth-order polynomial for $f_b$\footnote{We have checked that our fit is numerically stable with respect to higher-order terms in the response function; the fourth-order polynomial captures the best balance between fit coverage and computational efficiency.} .

Once $f_b$ is constructed for each bin in the distribution, all that remains is to define a goodness of fit function between theory and data, and minimise it to obtain exclusion contours for \{\co{i}\}.

\section{Improving the top EFT fit at the LHC}
\label{sec:improve}

\subsection{The impact of high $p_T$ top final states}

As noted in the introduction, the bounds obtained on top quark operators from early LHC data are rather weak. In principle, differential distributions provide much more sensitivity to higher-dimensional operators than inclusive rates, because they isolate the regions of phase space where the operators are most sensitive. Typically, however, the differential measurements used in the fit have been based on standard top reconstruction techniques, which, while providing good coverage of the low $p_T$ `threshold' region, suffer from poor statistical and systematic uncertainties in the tails of distributions, precisely the region of phase space we aim to isolate.

\begin{figure}[t!]
\begin{center}
     \includegraphics[width=0.47\textwidth]{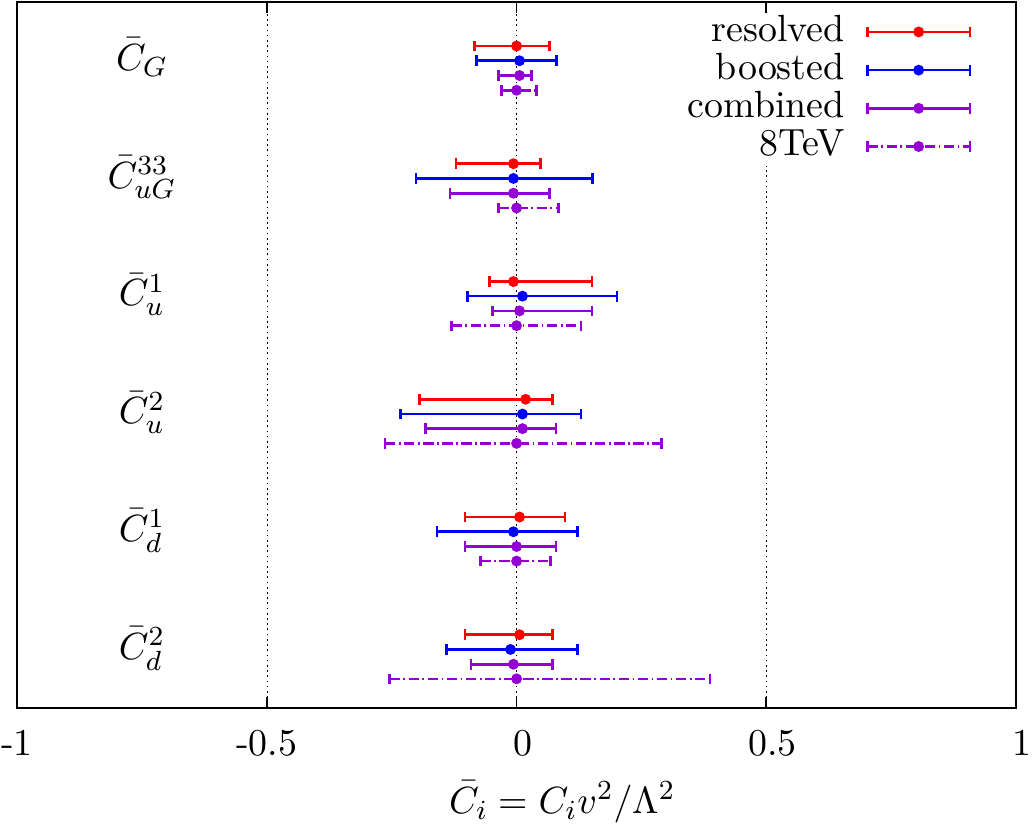}
         \caption{Individual 95\%~bounds on the operators considered here, from the boosted analysis and the resolved fat jet analysis, and the combined constraint from both, assuming 20\%~systematics and 30~\ifb of data. We also show existing constraints from unfolded 8 TeV $p_T$ distributions published in~\cite{Khachatryan:2015oqa} and~\cite{Aad:2015mbv}, showing the sizeable improvement even for a modest luminosity gain.} \label{fig:constraints}
  \end{center}
\end{figure}

Moreover, the measurements used were typically unfolded; that is, the final-state objects were corrected for detector effects and the actual measured `fiducial' cross section extrapolated to the full phase space, without cuts. This includes the treatment of reducible as well as irreducible backgrounds, which we implicitly understand as part of experimental systematic uncertainties in the following. Unfolded distributions substantially ease the workflow of our fit, since we can compare them directly to parton level quantities without the need for showering, hadronisation and detector simulation at each point in the parameter space. However, the extrapolation from the fiducial to full phase space, which makes use of comparing to Monte Carlo simulations, necessarily biases the unfolded distributions towards SM-like shapes. It also introduces additional correlations between neighbouring bins, broadening the $\chi^2$.

\setlength\extrarowheight{2.5pt}
\begin{table}[!b]
\begin{tabular}{ll@{\qquad}ll} \hline
\textit{Leptons} & $p_T > 30$ GeV & \\
			& $|\eta|  < 4.2$ \\ 
\textit{Missing energy} & $E_T^{\text{miss}} > 30$ GeV & \\
\textit{Small jets} & anti-$k_T$ $R = 0.4$ \\
			& $p_T > 30$ GeV , $|\eta| < 2 $ \\	
\textit{Fat jets} & anti-$k_T$ $R = 1.2$ \\
			& $p_T > 200$ GeV , $|\eta| < 2 $ \\	\hline
\textbf{Resolved} & $\geq$ 4 small jets w/$\geq$ 2 b-tags \\
\textbf{Boosted} & $\geq$ 1 fat jet, $\geq$ 1 small jet w/ b-tag \\	 \hline		
\end{tabular}
\caption{\label{tab:cuts} Summary of the physics object definitions and event selection criteria in our hadron-level analysis.}
\end{table}

\begin{figure*}[t!]
\begin{center}
     \includegraphics[width=0.9\textwidth]{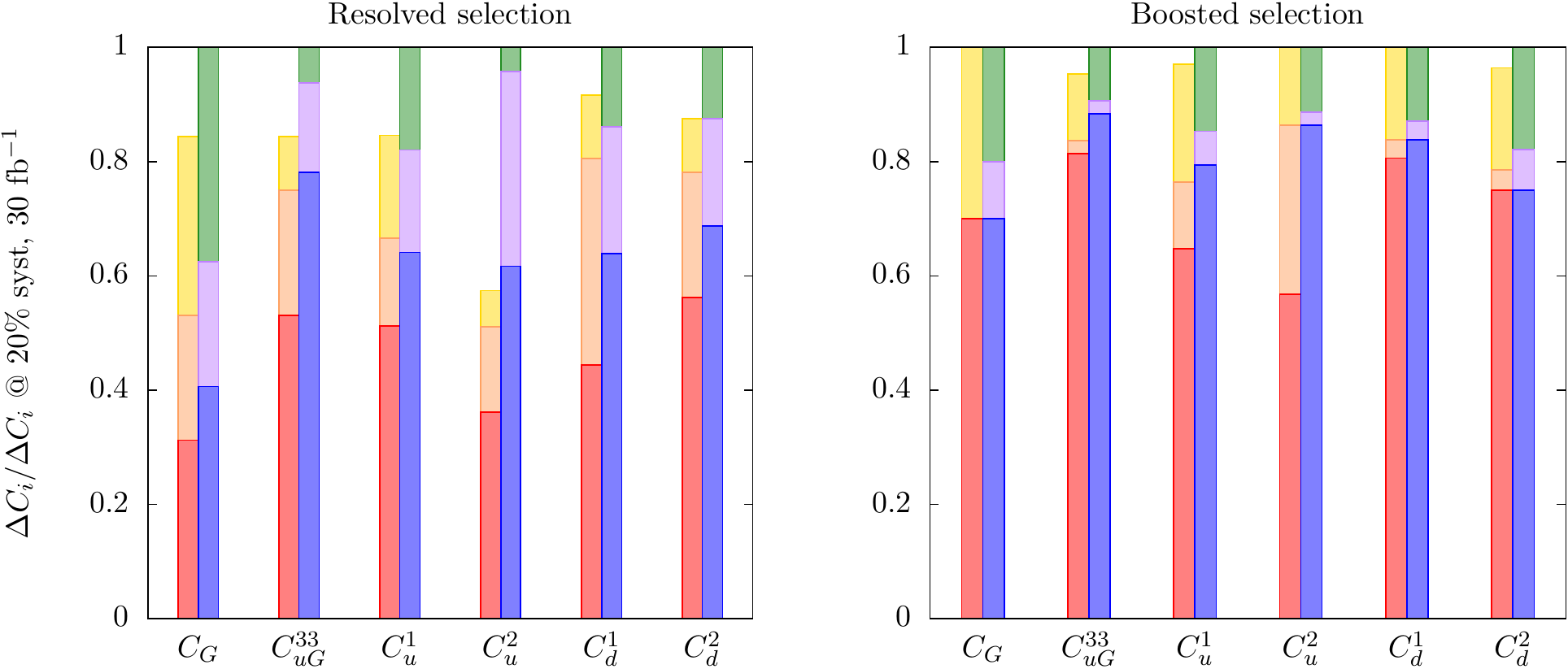}
         \caption{Fractional improvement on the 95\%~confidence intervals for the operators considered here, with various combinations of luminosity and experimental systematics considered. We take the width of the 95\% confidence limit obtained from 20~\% systematic uncertainty and 30~\ifb of data as a baseline (green bar), and normalise to this, i.e. we express constraints as a fractional improvement on this benchmark. The purple and blue bars represent respectively, 300~\ifb and 3~\iab of data, also at  20\% systematics, while the yellow, orange and red are the analogous data sample sizes for 10\% systematics. \label{fig:norm_constraints}}
  \end{center}
\end{figure*}

For top pair production, being a $2\to 2$ process, the relevant observables which span the partonic phase space are scattering angle and partonic centre-of-mass energy. All other observables are functions of these parameters, of which the top quark transverse momentum is the most crucial in determining the quality and efficiency of the boosted top tagging approach \cite{Plehn:2009rk,Plehn:2010st,Plehn:2011tg,Altheimer:2013yza,Schaetzel:2013vka,Plehn:2011sj,Backovic:2013bga} which we will employ in the following. The advantage of selecting high $p_T$ objects is thus twofold~\cite{Englert:2014oea}. Firstly, by making use of sophisticated reconstruction techniques for boosted objects, we move to the region of phase space where the effects of heavy new degrees of freedom will be most pronounced, as illustrated in Fig. \ref{fig:boosted_pt}, and secondly, jet substructure techniques require, by definition, a hadron-level analysis, so we avoid the model-dependence that fitting parton-level distributions to unfolded measurements suffers from. 

The sting in the tail for analyses selecting high $p_T$ objects is, of course, low rates. At 13 \tev, for instance, we find that 90\% of the cross section comes from the resolved region $p^t_T < 200$~\gev.\footnote{We choose $p^t_T \geq 200$~\gev as benchmark point of the boosted selection as the top tagging below this threshold suffers from large mistag rates and small efficiencies.} We thus aim to quantify at what stage in the LHC programme, if at all, the increased sensitivity in this region can compensate for the relatively poor statistics. Our analysis setup, as implemented in \textsc{Rivet} \cite{Buckley:2010ar}, is as follows (summarised in Tab.~\ref{tab:cuts}):

Restricting ourselves to the semileptonic top pair decay channel,  we first require a single charged lepton with $p_T > 30$ GeV\footnote{We do not consider $\tau$ decays here to avoid the more involved reconstruction.}, and find the $E_T^{\text{miss}}$ vector which we require to have a magnitude $ > 30$ GeV. The leptonic $W$-boson is reconstructed from these by assuming it was produced on-shell. Jets are then clustered using the anti-$k_T$ algorithm~\cite{Cacciari:2008gp} using {\textsc{FastJet}} \cite{Cacciari:2011ma} in two separate groups with $ R=(0.4,1.2) $ requiring $p_T >(30,200) $ \gev respectively, and jets which overlap with the charged lepton are removed. The $R=1.2$ fat jets are required to be within $|\eta| < 2$, and the $R=0.4$ small jets are b-tagged within the same $\eta$ range with an efficiency of $70\%$ and fake rate of $1\%$~\cite{ATLAS:2012ima}. 

If at least one fat jet and one b-tagged small jet which does not overlap with the leading fat jet exists, we perform a boosted top-tag of the leading fat jet using \textsc{HEPTopTagger}~\cite{Plehn:2009rk,Plehn:2010st,Kasieczka:2015jma} and reconstruct the leptonic top candidate using the leading, non-overlapping b-tagged small jet and the reconstructed leptonic $W$.

If no fat jet fulfilling all the criteria exists, we instead require at least 2 b-tagged small jets and 2 light small jets. If these exist we perform a resolved analysis by reconstructing the hadronic $W$-boson by finding the light small jet pair that best reconstructs the $W$ mass, and reconstruct the top candidates by similarly finding the pairs of reconstructed $W$-boson and b-tagged small jet that best reconstruct the top mass.

Finally, regardless of the approach used, we require both top candidates to have $|m_\text{cand} - m_\text{top}| < 40$ GeV. If this requirement is fulfilled the event passes the analysis.

\subsection{Results}

\subsubsection*{Impact of experimental precision}

Using a sample size of 30~\ifb with a flat 20\% systematic uncertainty (motivated by typical estimates from existing experimental analyses by ATLAS~\cite{Aad:2015hna} and CMS~\cite{Khachatryan:2016gxp}) on both selections as a first benchmark, the 1-dimensional 95\% confidence intervals on the operators considered here are presented in Fig.~\ref{fig:constraints}. All the bounds presented here are `one-at-a-time', i.e. we do not marginalise over the full operator set. Our purpose here is to highlight the relative contributions to the allowed confidence intervals here, rather than to present a global operator analysis.

As a general rule, the increased sensitivity to the Wilson coefficients offered by the boosted selection is overpowered by the large experimental systematic uncertainties in this region, and the combined limits are dominated by the resolved top quarks. The exception to this rule is the coefficient $C_G$ from the operator $O_G = f_{ABC} G^{\mu,A}_\nu G^{\nu,B}_\lambda G^{\lambda,C}_\mu.$ Expanding out the field strength tensors leads to vertices with up to six powers of momentum in the numerator, more than enough to overcome the na\"{i}ve $1/\hat{s}^2$ unitarity suppression. Large momentum transfer final states thus give stronger bounds on this coefficient, even with comparatively fewer events.

With these constraints as a baseline, it is then natural to ask by how much they can be improved upon when refinements to experimental precision are made. The constraints are presented in Fig.~\ref{fig:norm_constraints} for different combinations of systematic and statistical uncertainties. We take the width of the 95\% confidence interval in Fig.~\ref{fig:constraints} as our normalisation (the green bars), and express the fractional improvements on the limits that can be achieved relative to this baseline, for each operator. The right bars (green, purple, blue) represent 20\% systematic uncertainties with, respectively 30, 300 and 3~\iab of data. The left bars (yellow, orange, red) represent the same respective data sample sizes, but with 10\% systematic uncertainties. 

\begin{figure*}[!t]
\begin{center}
 \includegraphics[width=0.42\textwidth]{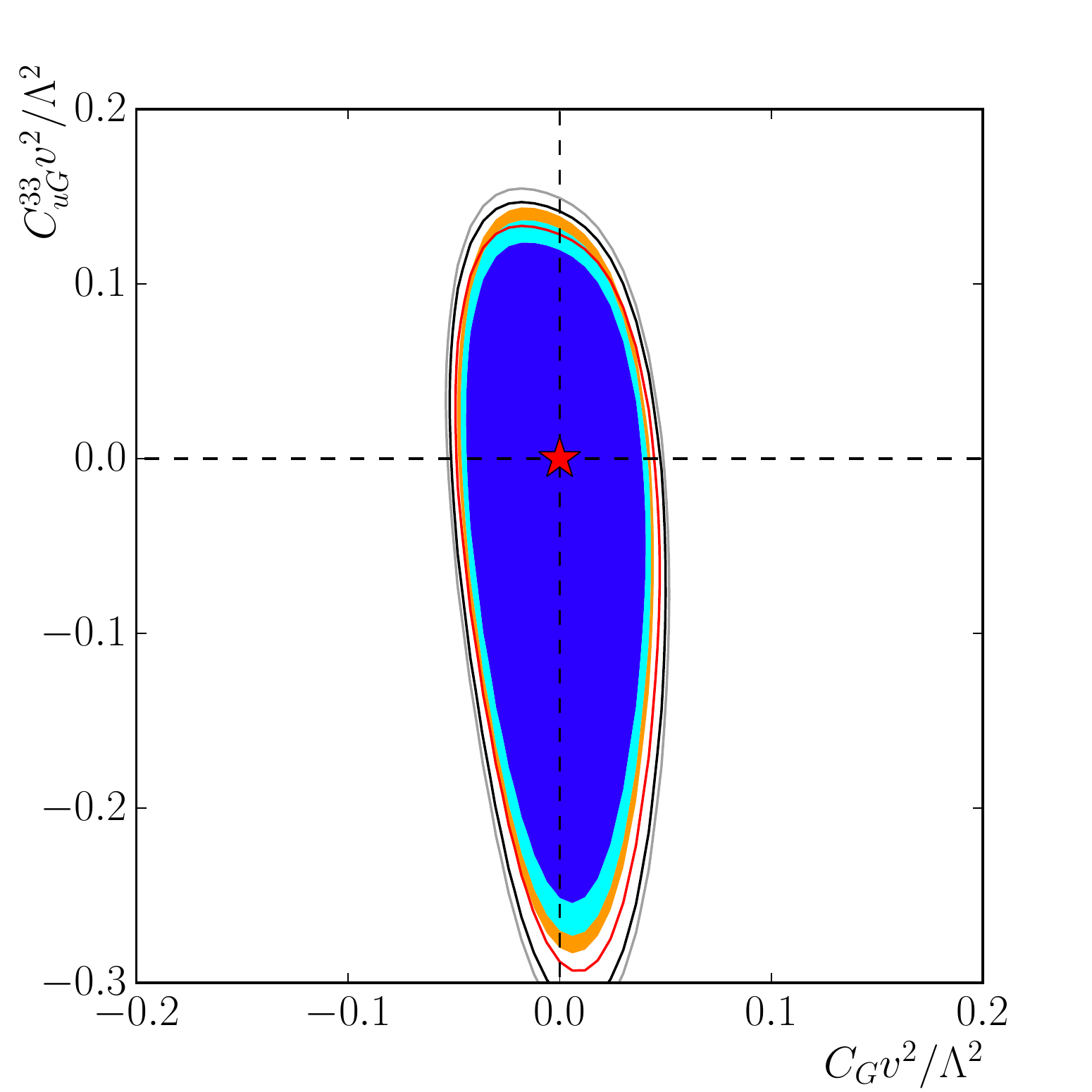}
  \hspace{1cm}
  \includegraphics[width=0.42\textwidth]{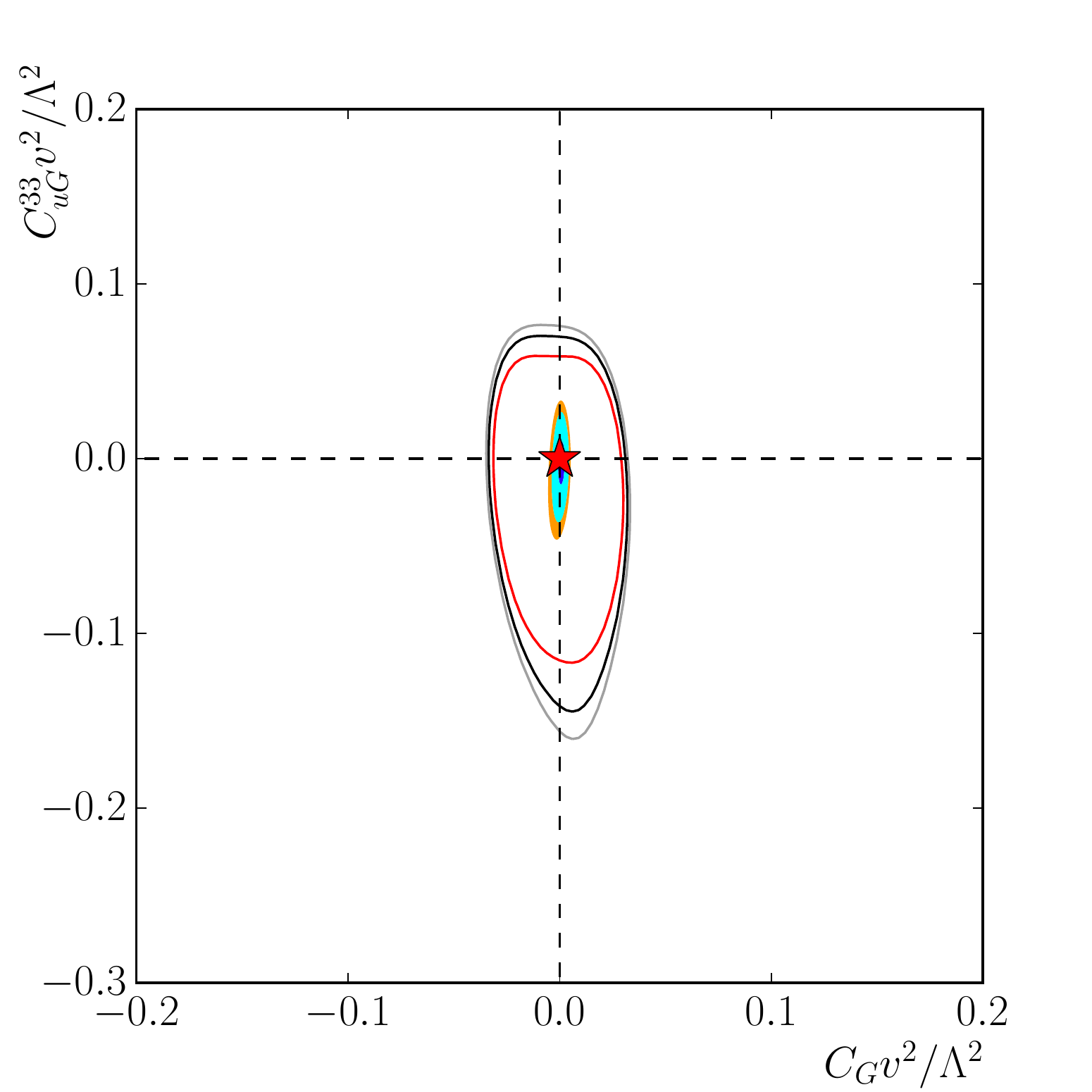}
  \caption{Left: 68\%, 95\% and 99\% confidence intervals for \co{G} and \co[33]{uG}, the lines are obtained using experimental (20\% systematics and 30~\ifb of data) uncertainties along with theoretical uncertainties, the filled contours using only experimental uncertainties. Right: the same plot, but using 10\% systematics and 3~\iab of data, showing the much stronger impact of theory uncertainties in this region.}
\label{fig:contours}
\end{center}
\end{figure*}

Beginning with the resolved selection, we find that the limits on the
coefficient $C_G$ can be improved by 40\% by going from 30~\ifb to
300~\ifb, and by a further 20\% when the full LHC projected data
sample is collected. Systematic uncertainties have a more modest
effect on this operator: at 3~\iab the limit on $C_G$ is only
marginally improved by a 10\% reduction in systematic
uncertainty. This merely reflects that $C_G$ mostly impacts the high
$p_T$ tail, so it can only be improved upon in the threshold region by
collecting enough data to overcome the lack of sensitivity. 8
TeV measurements are already constraining the relevant phase space
region efficiently and the expected improvement at 13 TeV is only mild
(see below).

For the chromomagnetic dipole operator $O^{33}_{uG}$, improving the experimental systematics plays much more of a role. A 10\% improvement in systematics, coupled with an increase in statistics from 30~\ifb to 300~\ifb leads to stronger limits that maintaining current systematics and collecting a full 3~\iab of data. Similar conclusions apply for the four-quark operators, to varying degrees, i.e. reducing systematic uncertainties can provide comparable improvements to collecting much larger data samples.

For the boosted selection, the situation is quite different. For all the operators we consider, improving systematic uncertainties by 10\% has virtually no effect on the improvement in the limits. This simply indicates that statistical uncertainties dominate the boosted region at 30~\ifb. For $C_G$, at 300~\ifb some improvement can be made if systematics are reduced, however we then see that systematic uncertainties saturate the sensitivity to $C_G$, i.e. there is no improvement to be made by collecting more data. For $C^{33}_{uG}$, a modest improvement can also be made both by reducing systematics by 10\% and by increasing the dataset to 300~\ifb. However, going beyond this, the improvement is minute. The four-quark operators again follow this trend, although $C^2_u$ shows much more of an improvement when going from 300~\ifb to 3~\iab.

\subsubsection*{The role of theoretical uncertainties}

The other key factor in the strength of our constraints is the uncertainties that arise from theoretical modelling. The scale and PDF variation procedure outlined in Sec. \ref{sec:elements} typically leads to uncertainties in the 10-15\% range. Fully differential $K$-factors for top pair production at NNLO QCD (i.e. to order $\mathcal{O}(\alpha_s^4)$) have become available, which have substantially reduced the scale uncertainties. The numbers quoted in Refs. \cite{Czakon:2015owf,Czakon:2016ckf} are for the Tevatron and 8 TeV LHC, and available only for the low to intermediate $p^t_T$ range ($p^t_T < 400$ \gev). Updated results for 13 TeV have become available only recently~\cite{Czakon:2016dgf}. It is worthwhile to ask what impact such an improvement could have on the constraints.

We put this question on a firm footing by showing in Fig. \ref{fig:contours} the 2D exclusion contours for the coefficients \co{G} and \co[33]{uG}, as obtained from combining the boosted and resolved limits, at fixed luminosity and experimental systematics, first using our NLO theory uncertainty, and also using \textit{no theory uncertainty at all}. For 30~\ifb the improvement is limited, indicating that at this stage in the LHC programme the main goal should be to first improve experimental reconstruction of the top quark pair final state. However, at 3~\iab the improvement is substantial, indicating that it will also become necessary to improve the theoretical modelling of this process, if the LHC is to augment its kinematic reach for non-resonant new physics. 

In addition to SM theoretical uncertainties, there are uncertainties relating to missing higher-order terms in the EFT expansion. Uncertainties due to to loop corrections and renormalisation-group flow of the operators \op[(6)]{i} are important for measurements at LEP-level precision~\cite{Berthier:2015gja,Berthier:2016tkq} where electroweak effects are also resolved. However, at the LHC we find them to be numerically insignificant compared to the sources of uncertainty that we study in detail here.  In addition, there is also the possibility of large effects due to dimension-8 operators, particularly owing to additional derivatives in the EFT expansion Eq.~\eqref{eqn:eftexpn}. Since the interference effects of omitted dimension-8 operators are formally of the same order as the retained quadratic terms in the dimension-6 operators, we emphasise that the numerical constraints presented here should be treated with caution. The only way to be certain that the omission of these terms is justified is to compute the effects of the interference of the relevant dimension-8 operators to a given process and demonstrate them to be small. This has been shown to be true for the $gg\to t\bar{t}$ subprocess\cite{Cho:1993eu,Cho:1994yu}. However, due to the large number of operators present there, this has not been studied for the $q\bar{q}\to t\bar{t}$ process. We leave a full computation of these effects as a future direction of study.

\subsection{Interpreting the results}

The whole purpose of the EFT approach is to serve as a bridge between the Standard Model and heavy degrees of freedom residing at some unknown mass scale $M_*$. Connecting the EFT to this scale, however, necessarily involves making assumptions about the couplings of this new physics. We can make statements about the relation between the constraints presented here and such a scale, however, by making general assumptions, such as perturbativity of the underlying new physics. 

Consider, for example, the simple case where the perturbative UV physics is characterised entirely by a single coupling $g_*$ and a unique mass scale $M_*$. Such a scenario could arise from integrating out a heavy, narrow resonance. In this case we have the simple tree-level matching condition
\begin{equation}
\frac{\co{i}}{\Lambda^2} = \frac{g_*^2}{M_*^2}.
\label{eqn:matching}
\end{equation}
Constraints on \co{i} then map onto allowed regions in the $g_*$-$M_*$ plane. In Fig. \ref{fig:eft_validity} we sketch these regions for illustrative values of \co{i}. In order for the EFT description of a given mass region to be valid, we must not resolve it our measurement. Therefore we impose a hard cut at $\sqrt{s} = 2$ TeV, obtained from the maximum $t\bar{t}$ invariant mass probed in our SM pseudodata. We also impose a generic perturbativity restriction $g_* \lesssim 4\pi$ to ensure that our EFT expansion is well-behaved and higher-dimensional operators do not affect the power counting.

We see that for large Wilson coefficients $\bco{i} \gtrsim 0.5 $ only a very small window of parameter space may be constrained, but the weak limits push the underlying coupling to such large values that loop corrections are likely to invalidate the simple relation of Eq. (\ref{eqn:matching}), making it hard to trust these limits. However, at 3 \iab, the projected constraints are typically $\bco{i} \lesssim 0.01 $, therefore, even for moderate values of the coupling $g_*$, our constraints are able to indirectly probe mass scales much higher than the kinematic reach of the LHC.

\begin{figure}[t!]
\begin{center}
     \includegraphics[width=0.5\textwidth]{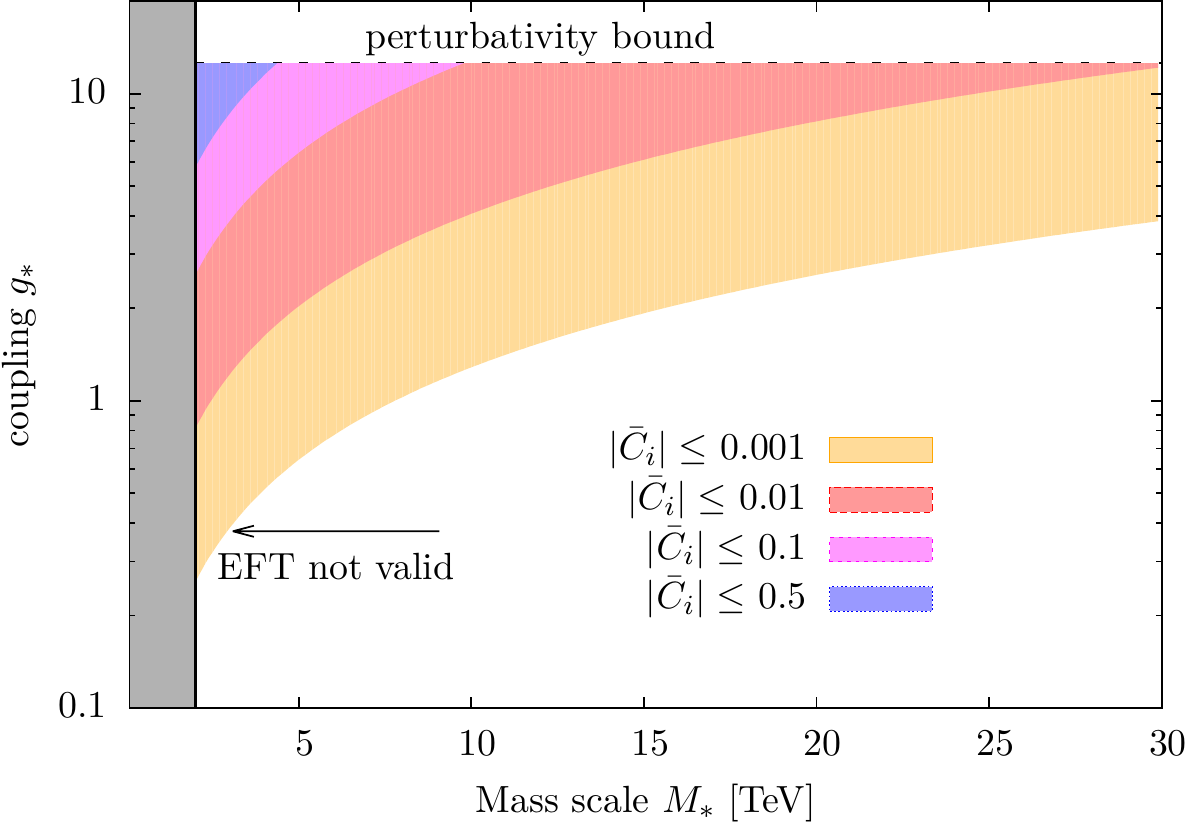}
         \caption{Areas in the new coupling-BSM mass scale plane (see also \cite{Englert:2014cva}), resulting from our fit coverage. Shaded areas are constrained in perturbative UV completions at a scale $M_\ast$, subject to the boundary condition Eq. (\ref{eqn:matching}). The shaded grey area is probed with by the pseudo-data of our fit. We do not consider unitarity bounds in this work.} 
         \label{fig:eft_validity}
  \end{center}
\end{figure}

\section{Summary and Conclusions}
\label{sec:conc}
The special role of the top quark in BSM scenarios highlights the importance of searches for new interactions in the top sector. Taking the lack of evidence of resonant new physics in the top sector at face value \cite{Aad:2012em,Aad:2012ans,Chatrchyan:2012yca}, we can assume that new interactions are suppressed by either weak couplings or large new physics scales. In both cases we can analyse the presence of new physics using effective field theory techniques. A crucial question that remains after the results from the LHC run 1 is in how far a global fit from direct search results will improve with higher statistics and larger kinematic coverage. We address this question focusing on the most abundant top physics-related channel $pp\to t\bar t$, which probes a relevant subset of top quark effective interactions. 
In particular, we focus on complementary techniques of fully-resolved vs. boosted techniques using jet-substructure technology, which are affected by different experimental systematic uncertainties. Sensitivity to new physics is a trade off between small statistical uncertainty and systematic control for low $p_T$ final states at small new physics-induced deviations from the SM expectation (tackled in fully-resolved analyses) and the qualitatively opposite situation at large $p_T$. For the typical parameter choices where top-tagging becomes relevant and including the relevant efficiencies, we can draw the following conclusions:
\begin{itemize}
\item Boosted top kinematics provide a sensitive probe of new interactions in $t\bar t$ production mediated by modified trilinear gluon couplings. In particular, this observation shows how differential distributions help in breaking degenerate directions in a global fit by capturing sensitivity in phenomenologically complementary phase space regions.
\item The sensitivity to all other operators detailed in Tab.~\ref{tab:ops} is quantitatively identical for boosted and fully-resolved analyses for our choice of $p_T^{\text{boost}}\geq 200~\text{GeV}$. Increasing the boosted selection to higher $p_T$ (where the top tagging will become more efficient) will quickly move sensitivity to new physics effects to the fully resolved part of the selection. The boosted selection is saturated by large statistical uncertainties for the for the typical run 2 luminosity expectation. These render systematic improvements of the boosted selection less important in comparison to the fully resolved selection, which provides an avenue to set most stringent constraint from improved experimental systematics. Similar observations have been made for boosted Higgs final states~\cite{Butterworth:2015bya} and are supported by the fact that the overflow bins in run 1 analyses provide little statistical pull~\cite{Buckley:2015lku}.
\item Theoretical uncertainties that are inherent to our approach are
  not the limiting factors of the described analysis in the forseeable
  future, but will become relevant when statistical uncertainties
  become negligible at very large integrated luminosity.
\end{itemize}

Boosted analyses are highly efficient tools in searches for resonant
new
physics~\cite{Joshi:2012pu,Aad:2012em,Aad:2012ans,Chatrchyan:2012yca}. Our
results show that similar conclusions do not hold for non-resonant new
physics effects when the degrees in questions do not fall inside the
kinematic coverage of the boosted selection anymore. Under these
circumstances, medium $p_T$ range configurations which maximise new
physics deviation relative to statistical and experimental as well as
theoretical uncertainty are the driving force in setting limits on
operators whose effects are dominated by interference with the SM
amplitude in the top sector. This also implies that giving up
  the boosted analysis in favor of a fully resolve analysis extending
  beyond $p_T^t\geq 200~\gev$ will not improve our results
  significantly. The relevant phase space region can be accessed with fully resolved techniques, with a large potential for improvement from the
experimental systematics point of view.

\acknowledgments
We would like to thank the members of our {\textsc{TopFitter}} collaboration. CE, KN, MR are grateful to the Mainz Institute for Theoretical Physics (MITP) for its hospitality and its partial support during the completion of this work. LM is thankful to the CERN TH Division for their affability and support. LM and MR are supported in part by the UK Science and Technology Facilities Council (STFC) under grant ST/L000446/1. KN is supported in part by the University of Glasgow College of Science \& Engineering through a PhD scholarship. LM, KN and MR are supported in part by the European Union as part of the FP7 Marie Curie Initial Training Network MCnetITN (PITN-GA-2012-315877).

\begin{widetext}
\appendix
\section{Numerical constraints}
For completeness we show the numerical values corresponding to the confidence intervals displayed in Fig. 2.
\begin{table*}[h!]
\setlength\extrarowheight{2.5pt}
\begin{tabular}{| c | c | c | c | c |} \hline 
 \textbf{Coefficient} & 		 \textbf{Resolved} & 		 \textbf{Boosted}  &		\textbf{Combined} &	\textbf{LHC8} \\ \hline  
$C_{G}v^2/\Lambda^2$ & 		($-$0.085,0.066) &  ($-$0.134,0.116)  & ($-$0.036,0.030)  & ($-$0.030,0.040) 	\\ \hline 
$C^{33}_{uG}v^2/\Lambda^2$ &  	($-$0.121,0.0484) & ($-$0.171,0.098)  &  ($-$0.133,0.066) & ($-$0.036,0.084)	\\ \hline 
$C^{1}_{u}v^2/\Lambda^2$ & 		 ($-$0.054,0.151) & 		 ($-$0.104,0.201) & 		($-$0.048,0.151) &  ($-$0.130,0.129)	\\ \hline 
$C^{2}_{u}v^2/\Lambda^2$ & 		 ($-$0.194,0.072) & 		 ($-$0.243,0.122) &  		($-$0.181,0.079) & 	($-$0.263,0.290) \\ \hline 
$C^{1}_{d}v^2/\Lambda^2$ & 		 ($-$0.102,0.0969) & 		 ($-$0.153,0.147) & 		($-$0.103,0.078) & ($-$0.072,0.068) \\ \hline 
$C^{2}_{d}v^2/\Lambda^2$ & 		($-$0.103,0.073) & 		 ($-$0.153,0.123) &          ($-$0.091,0.0727) & ($-$0.254,0.387)	 \\ \hline 
\end{tabular}
\caption{Numerical values of the 95\% confidence intervals presented in Fig.~\ref{fig:constraints}.} 
\end{table*}
\end{widetext}

\end{document}